\documentclass[10pt,preprint2]{aastex}

\slugcomment{(Submitted to ApJ)}

\shorttitle{The TeV spectrum of H\,1426+428}
\shortauthors{Petry et al.}

\begin{document}


\title{The TeV spectrum of H\,1426+428}


\author{D.~Petry\altaffilmark{1},
I.H.~Bond\altaffilmark{2}, 
S.M.~Bradbury\altaffilmark{2},
J.H.~Buckley\altaffilmark{3}, 
D.A.~Carter-Lewis\altaffilmark{1},  
W.~Cui\altaffilmark{4},  
C.~Duke\altaffilmark{5}, 
I.~de la Calle Perez\altaffilmark{2}, 
A.~Falcone\altaffilmark{4},
D.J.~Fegan\altaffilmark{6},
S.J.~Fegan\altaffilmark{7}\altaffilmark{,15}, 
J.P.~Finley\altaffilmark{4},
J.A.~Gaidos\altaffilmark{4}, 
K.~Gibbs\altaffilmark{7},
S.~Gammell\altaffilmark{6},
J.~Hall\altaffilmark{9},
T.A.~Hall\altaffilmark{1}\altaffilmark{,16},
A.M.~Hillas\altaffilmark{2}, 
J.~Holder\altaffilmark{2}, 
D.~Horan\altaffilmark{7},  
M.~Jordan\altaffilmark{3}, 
M.~Kertzman\altaffilmark{8}, 
D.~Kieda\altaffilmark{9},
J.~Kildea\altaffilmark{6}, 
J.~Knapp\altaffilmark{2},
K.~Kosack\altaffilmark{3},  
F.~Krennrich\altaffilmark{1}, 
S.~LeBohec\altaffilmark{1}, 
P.~Moriarty\altaffilmark{10},  
D.~M\"uller\altaffilmark{10},
T.N.~Nagai\altaffilmark{9},
R.~Ong\altaffilmark{11},   
M.~Page\altaffilmark{5},
R.~Pallassini\altaffilmark{2},
B.~Power-Mooney\altaffilmark{6},
J.~Quinn\altaffilmark{6},
N.W.~Reay\altaffilmark{12}, 
P.T.~Reynolds\altaffilmark{13},           
H.J.~Rose\altaffilmark{2}, 
M.~Schroedter\altaffilmark{7}\altaffilmark{,15},
G.H.~Sembroski\altaffilmark{4},  
R.~Sidwell\altaffilmark{12},  
N.~Stanton\altaffilmark{12},
S.P.~Swordy\altaffilmark{14}, 
V.V.~Vassiliev\altaffilmark{9},
S.P.~Wakely\altaffilmark{14},
G.~Walker\altaffilmark{9},  
T.C.~Weekes\altaffilmark{7} }


\altaffiltext{1}{Department of Physics and Astronomy, Iowa State
University, Ames, IA 50011}

\altaffiltext{2}{Department of Physics, University of Leeds,
Leeds, LS2 9JT, Yorkshire, England, UK}

\altaffiltext{3}{Department of Physics, Washington University, St.~Louis,
MO 63130}

\altaffiltext{4}{Department of Physics, Purdue University, West
Lafayette, IN 47907}

\altaffiltext{5}{Physics Department, Grinnell College, Grinnell, IA 50112}

\altaffiltext{6}{Physics Department, National University of Ireland,
Belfield, Dublin 4, Ireland}

\altaffiltext{7}{ Fred Lawrence Whipple Observatory, Harvard-Smithsonian
CfA, Amado, AZ 85645}

\altaffiltext{8}{Physics Department, De Pauw University, Greencastle, 
                   IN, 46135}

\altaffiltext{9}{High Energy Astrophysics Institute, University of Utah,
Salt Lake City, UT 84112}

\altaffiltext{10}{School of Science, Galway-Mayo Institute of Technology,
Galway, Ireland}

\altaffiltext{11}{Department of Physics, University of California, Los Angeles, CA 90095}

\altaffiltext{12}{Department of Physics, Kansas State University, Manhattan, KS 66506}

\altaffiltext{13}{Department of Physics, Cork Institute of Technology, Cork, Ireland}

\altaffiltext{14}{Enrico Fermi Institute, University of Chicago,  Chicago, IL 60637}

\altaffiltext{15}{Department of Physics, University of Arizona, Tucson, AZ 85721}

\altaffiltext{16}{Department of
Physics and Astronomy, University of Arkansas, Little Rock, AR 72204}


\begin{abstract}
The BL Lac object H1426+428 was recently detected as a high energy
$\gamma$-ray source by the VERITAS collaboration \citep{horan02}.
We have reanalyzed the 2001 portion of the data used in the detection
in order to examine the spectrum of H1426+428 above 250 GeV.
We find that the time-averaged spectrum agrees with a power law of the shape
$$
(\frac{\mathrm{d}F}{\mathrm{d}E})(E) = 
 10^{-7.31 \pm 0.15_{\mathrm{stat}} \pm 0.16_{\mathrm{syst}}}\cdot
 E^{{-3.50 \pm 0.35_{\mathrm{stat}}\pm 0.05_{\mathrm{syst}}}}
 \ \mathrm{m}^{-2}\mathrm{s}^{-1}\mathrm{TeV}^{-1}
$$
The statistical evidence from our data
for emission above 2.5 TeV is 2.6 $\sigma$.
With 95\,\% c.l., the integral flux of H1426+428 above 2.5 TeV
is larger than 3\% of the corresponding flux from the Crab Nebula.
The spectrum is consistent with the (non-contemporaneous) measurement 
by \citet{aharon02} both in shape and in normalization. Below 800 GeV, the data 
clearly favours a spectrum steeper than that of any other TeV Blazar observed so 
far indicating a difference in the processes involved either at the source or 
in the intervening space.
\end{abstract}



%% the last two numeral of the year of publication as o4ur KEY for

\section{Introduction}

The BL Lac object H1426+428 was discovered at optical wavelengths
with a redshift of 0.129 by \citet{remil89}. Measurements 
of the spectral energy distribution  of this close object 
are still sparse and do not sufficiently cover the expected broad
two-peak structure that is
familiar from other BL Lac objects (see, e.g., \citet{donato01}).
The first peak is expected at X-ray energies and is thought to represent 
the synchrotron emission from relativistic electrons in the source. 
The second peak is expected at $\gamma$-ray energies and is explained 
in  so-called leptonic models as stemming from low-energy photons 
which are inverse-Compton scattered to $\gamma$-ray energies by the 
same population of relativistic electrons that causes the synchrotron 
radiation. Alternative models
attribute the second peak and also part of the X-ray emission
to processes involving protons which are co-accelerated with the X-ray-emitting 
electrons or even partially produce them.  For recent reviews see, e.g., 
\citet{ghisell00}, \citet{rachen00}, \citet{sikora00}.  

 A lower limit on the position 
of the X-ray peak of H1426+428 was placed by \citet{costa01} at 100 keV.
Currently the only published measurements above 100 keV
come from Cherenkov telescopes above 280 GeV \citep{horan02} and
above 700 GeV \citep{aharon02}.  \citet{costa02} predict the peak of the
$\gamma$-ray emission of H1426+428 to be at several ten GeV.

The study of the high-energy spectrum of this ``new'' TeV source
is especially interesting since its redshift is four times as large as
that of Mkn 421 and Mkn 501, the only other extragalactic objects
detected at TeV energies with good spectral information. Due to this
larger distance, it is expected that
signs of absorption of the $\gamma$-radiation via interaction with the 
intergalactic optical and infra-red background (IIRB) will be more pronounced,
possibly permitting one to infer constraints on the IIRB photon density.

In this letter, we present the results of a spectral reanalysis of 
H1426+428 observations made with the Whipple 10 meter $\gamma$-ray 
telescope  on Mt. Hopkins, Arizona, in the first half of 2001. 
We use our standard method described 
in \citet{mohanty98} for deriving the spectrum.
Due to the weakness of the source and the special interest
in the emission at the highest energies, we then examine the
emission above 2.5 TeV with specially developed cuts.

\section{Dataset}

\begin{table}[h]
\caption{\label{tab-gen} General properties of the 
H1426 dataset and the corresponding OFF dataset.}
\begin{center}
\begin{tabular}{lcc}
    & ON & OFF \\
\tableline
first MJD & 51940 & 51875 \\
last MJD & 52275 & 52275 \\
raw number of events & 3698176 & 3826088 \\
min. zenith angle & 10$^\circ$  &  6$^\circ$ \\
max. zenith angle & 38$^\circ$  &  38$^\circ$ \\
mean zenith angle & 18.0$^\circ$ & 18.7$^\circ$ \\
total observation time  & 38.1 h  & 39.9 h  \\
\tableline
\end{tabular}
\end{center}
\end{table}

The detection of H1426+428 (H1426 in the following) as presented in
\citet{horan02}  is based on four separate datasets from
1995-98, 1999, 2000 and 2001 respectively. Since the sensitivity
of the Whipple telescope was considerably improved over from 1995 to 2001
and the largest fraction of the observation time was invested
in 2001, it was only in the 2001 dataset that
a $\gamma$-ray signal with significance $>5 \sigma$ was found. 
For this observing period, contemporaneous observations of
the Crab Nebula are available \citep{krenn01} containing high-significance
$\gamma$-ray signals which can serve to calibrate the telescope. 
In order to derive a spectrum for H1426, we have therefore  chosen to 
confine ourselves to the 2001 dataset.

The dataset used here consists of 87 pairs of on-source (ON) and off-source (OFF)
observations (``runs''). The run duration for the ON data varied between
10 and 28 minutes resulting in a total exposure time of 38.1 hours.
 The OFF dataset serves to determine the background 
caused by hadronic cosmic rays. 
Out of the 87 ON-OFF pairs in our dataset,
34 were taken in actual ON-OFF mode, i.e. with exactly
matching zenith angle distribution and night sky background
conditions.
The OFF data for the remaining 46 ON runs was obtained by selecting  
matching OFF runs out of the pool of other OFF data taken in early 2001.
The matching was based on how close to each other the runs were in terms of
date, weather conditions, elevation and night sky brightness.
This procedure results in a good agreement of the zenith angle distribution,
but small discrepancies remain. Also, some of the OFF runs have a longer
duration than their ON partners. As a result, the total OFF exposure time
is 4.7 \% longer than the ON exposure time. These differences were eliminated
in later stages of the analysis by using the number of events not coming
from the source direction to scale the background rate. This scaled background
should be equal for both the ON and the OFF data.

\section{Data analysis}

\subsection{Determination of image parameters}

The determination of the image parameters WIDTH, LENGTH, ALPHA, SIZE etc.
for ON and OFF data was 
carried out following the standard procedure described in
\citet{reynolds93} and \citet{cawley93}.

\subsection{Monte Carlo simulation}

Our data was taken between 10$^\circ$ and 38$^\circ$ zenith angle with
less than 10\,\% taken at zenith angles between 30$^\circ$ and 38$^\circ$.
The average zenith angle is near 20$^\circ$. The properties of 
air showers change as a function of the cosine of the zenith angle.
Given the relatively large statistical error of the gamma signal from H1426 
(see below), adequate accuracy for the determination of the detector sensitivity 
can therefore already be reached by simulating air showers only at the
average zenith angle.

The shower simulation code ISUSIM (based on ``kascade'' by \citet{kertz94},
see \citet{mohanty98}) was used to generate $3.8\times10^6$
showers at zenith angle 20$^\circ$ between 0.05\,TeV and 100\,TeV 
with a spectral index of 2.5
(MC set 1). Other spectral indices were simulated by re-weighting.
The simulation of the Whipple telescope was carried out with
the ``simveritas'' code (Duke \& LeBohec 2002, VERITAS internal report).
The telescope parameters in
this simulation were identical with those used in \citet{krenn01}.

\subsection{Derivation of the ``Extended Cuts''}

\label{sec-cuts}

Using MC set 1 and the spectrum analysis algorithms  developed in \citet{mohanty98},
the ``extended cuts'' for the $\gamma$-hadron separation in our dataset were 
derived. These utilize only the image parameters WIDTH, LENGTH and ALPHA.  
They are derived
as a second-order polynomial function of log(SIZE) by approximating
the distributions of WIDTH and LENGTH and ALPHA as Gaussians and
determining the mean $a$ and the standard deviation $\sigma$ as a function of
log(SIZE). The lower cuts on WIDTH and LENGTH respectively are then of 
the form $(a - t \sigma)$ while the upper cuts are of the form $(a + t \sigma)$.
The upper cut on ALPHA is also of the latter form.
The constant $t$, the ``tolerance'', determines the efficiency of the cuts.
It is generally chosen to be 2.0 in order to minimize possible systematic effects 
arrising from imperfections of the Monte Carlo simulations. 
In addition to the extended cuts, standard quality cuts were imposed as
in \citet{krenn01}.

\begin{figure}
\plotone{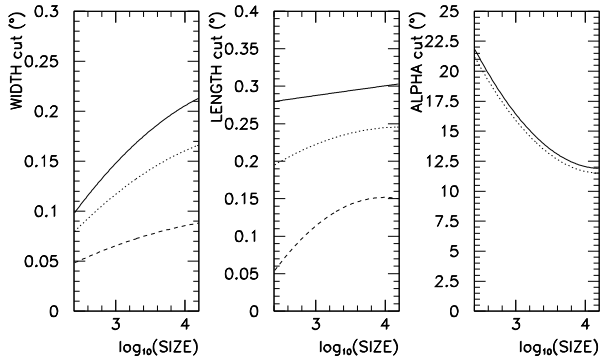}
\caption{\label{fig-excuts} The upper (solid) and lower (dashed) ``extended cuts'' 
on the image parameters WIDTH, LENGTH and ALPHA as a function of log(SIZE) with tolerance $t=2.0$. These cuts minimize systematic errors stemming from the simulations.
The dotted lines show the upper cuts for maximum significance above 2.5~TeV
(optimized on Crab Nebula data, see section \protect\ref{sec-above2.5}).} 
\end{figure}

\subsection{Background determination}

\label{sec-norm}

\begin{figure}
\plotone{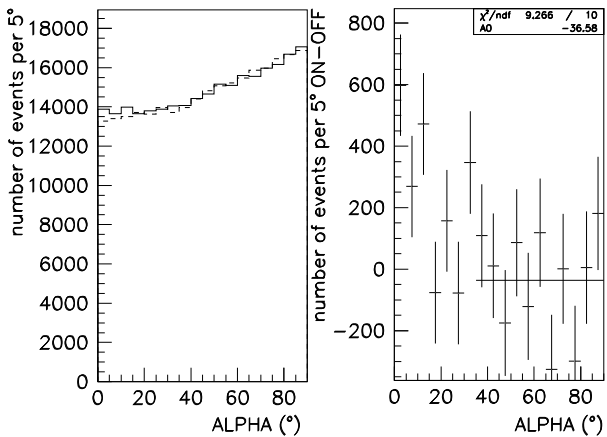}
\caption{\label{fig-alpha2} {\bf (a)} (left) ALPHA distribution
for the H1426 dataset ON (solid) and OFF (dashed, scaled by a factor 0.9062) after 
application of the extended cuts with $t=2.0$ described in section \protect\ref{sec-cuts} except the cut
on ALPHA. {\bf(b)} (right) the difference between the two
distributions shown in (a) with a constant function fitted to
the region ALPHA$>30^\circ$.}
\end{figure}

Due to the 5\,\% difference in exposure time and the remaining differences 
in elevation and night sky background, the number of events in the region 
ALPHA$>30^\circ$ (where according to simulations
there should be no significant contribution caused by $\gamma$-rays
from the source) is smaller for the ON data (184440 events) than 
it is for the OFF data (203531 events). The ratio $r$ of these numbers
(here 0.9062) can be used to scale the number of OFF events at small
ALPHA in order to determine the background.
Figure \ref{fig-alpha2}a shows the scaled OFF data distribution
(with $r = 0.9062$) superimposed on the ON distribution.

After applying the extended ALPHA cut (see Figure \ref{fig-excuts})
the number of events is 51819 ON and 50279 OFF with an excess of
1540 events and significance 5 $\sigma$.
Figure \ref{fig-alpha2}b shows that the shape of the ON and the OFF distributions
agrees well in the normalization region (ALPHA$>30^\circ$).

\subsection{Energy reconstruction and binning}

\label{sec-energy}

An algorithm described in \citet{mohanty98} was used to derive the coefficients in an analytical
expression for the estimated
primary $\gamma$-ray energy $E_{\mathrm{est}}$ from the 
Monte Carlo dataset 1 (see above). This expression is of
the form
$$
\log_{10}(E_{\mathrm{est}}) = a_1 + a_2 L + a_3 D + a_4 L^2 
       + a_5 D^2 + a_6 D L
$$
where $L = \log_{10}(\mathrm{SIZE})$, $D = \mathrm{DIST}$.
This estimator gives a roughly constant energy resolution above 300 GeV.
The energy resolution (the standard deviation in $\log_{10}(E_{\mathrm{est}})$) was 
determined near 400\,GeV to be 0.166, i.e. $\Delta E/E$ = 33\,\% . 
This value is used to determine the binning
of the data with respect to  $\log_{10}(E_{\mathrm{est}})$. Statistics theory 
\citep{scott79} suggests that the ideal compromise between maximal structure 
resolution and
minimal error in spectral parameters is achieved for weak signals if the energy
bin width is chosen to be two standard deviations. In our case 
we must choose the energy bin width equal to $2\times0.166 = 0.332$.

The collection area of the telescope after our cuts on the image parameters
reaches 10\,\% of the maximum value (75,000 \,m$^2$) at an energy
of 250\,GeV. The peak $\gamma$-rate for a standard
spectrum of index 2.7 is observed near 410 GeV.
We choose 250 GeV to be the lower edge of the second energy bin. 
The first energy bin will be excluded from fits of spectral functions. 
We find this necessary
since the
systematic errors of the simulation at the trigger 
threshold are much larger than at higher energies. 
The first column  of table \ref{tab-eventsebint2} gives the logarithmic centers 
of the equidistant energy bins which result from these choices.

In order to ensure that the normalization constant $r$ is independent of
$E_{\mathrm{est}}$, the ratio was measured separately in each energy bin.
We find that the estimated energies
in the H1426 ON dataset are systematically smaller than those in the OFF dataset
by a small constant factor.
By varying the factor until the scaling ratio $r$ is 
the same within the statistical errors in all energy bins,
we determine the value of $K$, the necessary correction factor for the
estimated energies in the OFF data,
to be $K = 0.9784 \pm 0.005$.

This small effect and its correction were studied in detail with datasets from observations
of the Crab Nebula and Mkn 421 (two strong sources) and was found to be
a result of the differences in the nightsky background conditions
and zenith angle distributions. It was verified that the application of 
the correction factor $K$ to the estimated energies of the OFF dataset 
results in correct spectral measurements. The statistical error of $K$
leads to an increase of the statistical errors of the spectral parameters
as discussed below.

\subsection{Derivation of the H1426 spectrum}

After application of the correction to the estimated energy and the scaling
to the OFF data, we bin
both ON and OFF events after all cuts in $\log_{10}(E_{\mathrm{est}})$.
For each of the energy bins, we determine a modified collection
area $A_{\alpha}$. Its value depends on the assumed spectral index $\alpha$ since
the collection area has to be averaged over a finite energy bin width and
the effects of the finite energy resolution will lead to spill-overs mostly
from lower to higher bins, especially if the spectrum is steep. 
The differential flux in the energy bin $i$ is then determined by
$$
 (\frac{\mathrm{d}F}{\mathrm{d}E})_i = \frac{\mathrm{ON}_i - \mathrm{OFF}_i}{A_{\alpha,i} T \Delta E_i}\cdot k
$$
where $T$ is the observation time and $\Delta E_i$ is the energy bin width.
The factor $k$ corrects for the approximate calculation of the
energy derivative over a finite bin width. It has for our case the value 0.9313 .
Table \ref{tab-eventsebint2} shows the event statistics for each energy bin.

\begin{table}[h]
\caption{\label{tab-eventsebint2} The event statistics for the 2001 H1426+428 dataset after application
of all extended cuts (tolerance $t=2.0$).
}
\begin{center}
{\small
\setlength{\tabcolsep}{3pt}
\begin{tabular}{lrrrclc}
 Energy  &\multicolumn{1}{c}{ON}  & \multicolumn{1}{c}{OFF} &\multicolumn{3}{c}{ON-OFF} & S \\
 (TeV) & (events) & (events) & \multicolumn{3}{c }{(events)} & ($\sigma$) \\
\tableline
 0.171 & 10341 & 10454 & -113 &$\pm$& 144 & -0.78  \\
 0.369 & 23149 & 22255 &  894 &$\pm$& 213 &  4.2 \\
 0.794 & 13887 & 13294 &  593 &$\pm$& 164 &  3.6 \\
 1.71 &   3424 &  3283 &  141 &$\pm$&  81 &  1.7 \\
 3.69 &    776 &   760 &   16 &$\pm$&  39 &  0.41 \\
 7.94 &    209 &   200 &    9 &$\pm$&  20 &  0.46 \\
 17.1 &     33 &    33 &    0 &$\pm$&   8 &  0.0 \\
\tableline
\end{tabular}
}
\end{center}
\end{table}

Starting with the assumption that the spectral index will be similar to 2.5,
we calculate the differential fluxes in each energy bin using $A_{2.5}$ and 
fit a power law to the resulting values excluding the bin 1 with its large
systematic errors. After iterating twice, using every time the spectral index
$\alpha$ resulting from the fit to recalculate $A_{\alpha}$ and then recalculating
the differential fluxes and repeating the fit, we reach convergence at 
 $\alpha = 3.50 \pm 0.30$.

If we vary the correction $K$ of the estimated energy for the OFF data within
its statistical errors, we find that the resulting $\alpha$ varies by $\pm0.18$.
This has to be regarded as an additional, to a good approximation
uncorrelated, error since the uncertainty of $K$
stems from a finite number of events which are different from those events
used to measure $\alpha$. The total statistical error on $\alpha$ is therefore
$\sqrt{0.30^2 + 0.18^2} = 0.35$. The systematic error on $\alpha$ for our method 
has to be determined from a dataset with small statistical errors. This was done
in \citet{krenn01}. By varying the cut tolerance $t$ between 1.5 and 3.0
on a Crab Nebula dataset, the systematic error was determined to be $\pm 0.05$.

The statistical error on the exponent of the flux constant is determined 
correspondingly $\sqrt{0.11^2 + 0.10^2} = 0.15$ while the systematic error is 
derived by varying the absolute energy calibration by $\pm$\,15\%. Note that
the steepness of the spectrum makes this systematic error especially large.
The final result is

\begin{tabular}{rcl}
$\frac{\mathrm{d}F}{\mathrm{d}E}$ & $=$ & $ 10^{-7.31 \pm 0.15_{\mathrm{stat}} \pm 0.16_{\mathrm{syst}}}$\\
& & $ \times E^{{-3.50 \pm 0.35_{\mathrm{stat}}\pm 0.05_{\mathrm{syst}}}}
 \ \mathrm{m}^{-2}\mathrm{s}^{-1}\mathrm{TeV}^{-1}$\\
\end{tabular}

\subsection{Evidence for emission above 2.5 TeV}
\label{sec-above2.5}

\begin{figure}[h]
\plotone{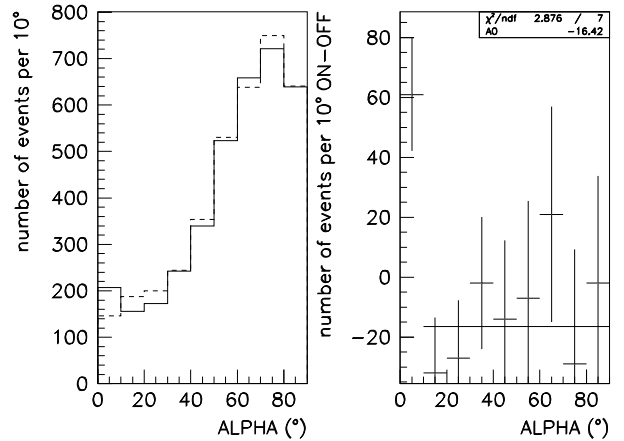}
\caption{\label{fig-above2.5} 
As Figure \protect\ref{fig-alpha2}, but only showing events with estimated energy
above 2.5 TeV and using tighter cuts optimized on Crab
Nebula data to give maximum significance above 2.5 TeV. The excess is expected
at ALPHA$<12^\circ$.}
\end{figure}

The extended cuts are designed to minimize systematic errors of the flux measurement.
When purely statistical evidence for the presence of emission is required, these
cuts are inadequate for weak sources. This is the case for H1426 above 2.5 TeV,
where the statistical significance of the excess obtained with the extended cuts
drops well below 1 $\sigma$ in each energy bin. The question whether emission
is present above a few TeV is, however, so important for this source that a special 
effort is justified.

Using 15.1 hours of contemporaneous Crab Nebula
observations and a matching OFF dataset, we optimize a set of cuts to give maximum 
significance
above 2.5 TeV. With the standard extended cuts, the significance is 5.5$\sigma$.
In the optimization,
we vary independently the tolerance $t$ for the upper cuts on WIDTH, LENGTH and ALPHA
keeping the tolerance for the lower cuts on WIDTH and LENGTH constant at $2.0$.
In addition we require that at least 70~\% of the excess obtained with the standard
extended cuts is retained.
The maximum significance (8.3\,$\sigma$) for the excess above an estimated energy
of 2.5 TeV is reached for tolerance $t=0.5$ for the upper WIDTH and LENGTH cuts
and $t=1.9$ for the ALPHA cut. These cuts are shown as dotted lines in Figure 
\ref{fig-excuts}.

When applied to the H1426 dataset (using a recalculated energy estimation
and correction $K$, see above), the optimized cuts improve the significance 
above 2.5 TeV as one would expect in the case that emission is actually present. 
Taking into account the error of the correction factor $K$
for the estimated energy of the background events, there are $59\pm22.7$ excess events
with estimated energy above 2.5 TeV. This corresponds to (12$\pm$5)\,\% of the excess
rate observed from the Crab Nebula. The statistical evidence for emission from H1426
above 2.5 TeV is 2.6 $\sigma$. Figure \ref{fig-above2.5} shows the ALPHA distribution
of the H1426 events with estimated energy above 2.5 TeV after the optimized cuts.

\section{Discussion}

\begin{figure}
\plotone{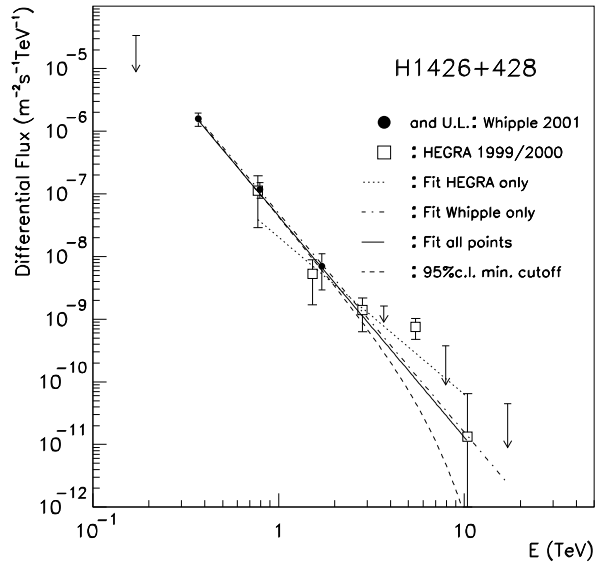}
\caption{\label{fig-finalspec} The differential energy spectrum of H1426+428
as measured in this analysis (filled points and upper limits) in 2001  and by HEGRA 
(open squares)
in 1999/2000 (Aharonian et al. 2002). The 84\% confidence level upper limits were 
obtained using the
 method by \protect\cite{helene83}. The solid line is the result of a power law fit
to all points except the upper limit at the lowest energy.  The dashed line shows
the scenario of the latter power law with an additional abrupt (super-exponential)
cutoff still consistent with all data at the 95\,\% confidence level. The minimum
cutoff energy found for this scenario is 6 TeV. Also shown is the power-law fit to 
the HEGRA points only (dotted) and the power-law fit to the Whipple points only 
(dot-dashed).
}
\end{figure}

The spectral index found in this analysis agrees well
with the first result ($\alpha = 3.55\pm0.5$)
published in \citet{horan02}.
The flux normalization constant is larger than that in \citet{horan02} but
well within the statistical and systematic errors of this result
which was obtained using a different analysis method and different
Monte Carlo data.

The only other published detection of H1426+428 at $\gamma$-ray energies comes
from HEGRA \citep{aharon02} and is based on a dataset of
similar size. The two results are compared in the following.

From a total of 44.4 hours HEGRA obtains a total number of excess events of 199.2
after loose cuts for their spectral analysis. The significance is 4.3 $\sigma$.
We obtain a slightly higher significance  and an excess of 1540 events 
from 38.1 hours of data using similarly loose cuts. 
The relative numbers of events recorded by HEGRA and Whipple are consistent
with the difference in energy threshold which is a factor of $\approx 2.5$ .
An estimation of the integral spectral index
from these numbers gives a value of 2.2 which already hints that the spectrum
must be steep if there was no significant change in the overall state
of the source between 1999/2000 and 2001. 

Direct comparison of the individual spectral points (Figure \ref{fig-finalspec})
shows a very good agreement indicating that there really was little change in the
overall state of the source between the observing periods.
Given the 30\% systematic errors of the absolute flux calibration of
both our and the HEGRA measurement,
this is consistent with the fact that the $\gamma$-ray rates measured by
\citet{horan02} for the 2000 and the 2001 H1426 dataset differ only by a 
factor $1.5$.

To make use of the total available information, we perform 
a fit of a power-law to the combined points from HEGRA and our analysis.
No adjustments to the overall normalization of any of the datasets was
performed since the points near 790 GeV agree very well.
The errors on the Whipple points included in this fit take into account
both the Gaussian error of the number of excess events 
($\sqrt{N_{\mathrm{on}}+ N_{\mathrm{off}}}$) and the uncertainty
stemming from the energy estimation for the background.
The fit to all points (still excluding the upper limit at the lowest energies) results in
a spectrum which is essentially identical  with what we obtain from our
points only. However, the statistical errors are smaller:
$$
 \frac{\mathrm{d}F}{\mathrm{d}E} = 
 10^{-7.36 \pm 0.07_{\mathrm{stat}}}\cdot
 E^{{-3.54 \pm 0.27_{\mathrm{stat}}}}
 \ \mathrm{m}^{-2}\mathrm{s}^{-1}\mathrm{TeV}^{-1}
$$
The reduced $\chi^2$ of this fit to 11 points is 0.94 . The measurements
are consistent with the assumption that the spectrum of
H1426 is a continuous powerlaw between 250 GeV and 17 TeV.

In order to further quantify the evidence for high-energy emission, we introduce
a super-exponential cutoff in the measured power-law. This cutoff can
be described analytically ($\exp(-(E/E_0)^2)$) and is resonably abrupt
such that it can serve as a parameter describing the energy above which
there is no emission from the source. By keeping the nominal values of
the power-law fit for normalization and spectral index and reducing
the cutoff energy $E_0$ until the $\chi^2$ of a fit to all significant points
(i.e. excluding the points shown in  Figure \ref{fig-finalspec} as upper limits)
has increased to 14.1 (95\% confidence level for 7 degrees of freedom)
we obtain $E_0 = 5.5$\,TeV.

Independently, we verify if this value for $E_0$ is consistent
with the fact that we observe from H1426 \  $12\pm5$\,\% of the
Crab Nebula $\gamma$-rate above 2.5 TeV (see section \ref{sec-above2.5}).
Taking into account the larger spill-over effects caused by the finite
energy resolution of our instrument and the steep spectrum of H1426
and furthermore the 15 \% uncertainty of our absolute energy calibration,
we find that a 95\,\% confidence level lower limit on the integral flux
of H1426 above 2.5 TeV can be put at 3\,\% of the Crab Nebula flux.
Using our measurement of the Crab Nebula flux with the same instrument, this
corresponds to
$$
	F_{\mathrm{H1426}}(E > 2.5 \mathrm{TeV}) > 1.06 \times 10^{-9} \mathrm{m}^{-2}\mathrm{s}^{-1} 
         \ \   (95\,\%\, \mathrm{c.l.})
$$
Varying the cutoff energy $E_0$ introduced above until the integral flux 
above 2.5 TeV has the value of this lower limit, we find a value of $E_0 = 6$\,TeV
which independently confirms the result from above. The spectrum with
$E_0 = 6$\,TeV is shown as a dashed line in Figure \ref{fig-finalspec}.

The HEGRA collaboration
attempts to reconstruct the intrinsic energy spectrum of the source by
assuming a  model for the intergalactic IR background and calculating
the expected absorption for $\gamma$-rays coming from H1426 (redshift 0.129). 
They arrive at an intrinsic spectrum
with index 1.9 and find that the ``upturn'' in the spectrum
around a few TeV may be explained by a decrease of the slope of the $\gamma$-ray
absorption as a function of energy.
Our data is not inconsistent 
with the absorbed spectrum derived by HEGRA. 
In fact, our lowest point at 370 GeV agrees very well
with the hypothesis of the absorbed spectrum, but not
with the extrapolation of the alternatively fitted 
power law (dotted line in Figure \ref{fig-finalspec}).
The latter can be excluded with a confidence level
$> 99.5$\,\% which argues for a flattening
of the spectrum at a few TeV. More data is needed to
confirm this result.

Between 250 GeV and 800 GeV, our data clearly favours a spectrum
which is steeper than that of any other
known TeV blazar at these energies. Since H1426 is also
the most distant known TeV blazar, the steepness of the
spectrum may be interpreted
as evidence for gamma-ray absorption in the intergalactic 
medium. However, the sample of TeV blazars still needs to be
enlarged and the intrinsic spectrum well separated from
absorption effects before one can come to firm conclusions.

\acknowledgments

The VERITAS Collaboration is supported by the U.S. Department of Energy,
National Science Foundation, the Smithsonian Institution, P.P.A.R.C. (U.K.) and
Enterprise Ireland.

\end{document}